\documentstyle[12pt]{elsart}
\def\be{\begin{equation}}
\def\ee{\end{equation}}
\newcommand\cref[1]{(\ref{#1})}

\begin{document}

\begin{frontmatter}
\title{An Efficient Implementation of \\ Flux Formulae
in Multidimensional \\ Relativistic Hydrodynamical Codes}
\author[Valencia]{M. A. Aloy \thanksref{thank}}
\author[Valencia]{J. A. Pons \thanksref{thank}}
\author[Valencia]{J. M$^{{\mbox a}}$ Ib\'a\~{n}ez \thanksref{thank}}
 
\address[Valencia] 
{Departament d'Astronomia i Astrof\'{\i}sica \\
Universitat de Val\`encia, 46100 Burjassot, Val\`encia, Spain}

\begin{abstract}
  We derive and analyze a simplified formulation of the 
numerical viscosity terms appearing in the expression of the 
numerical fluxes associated to several 
High-Resolution Shock-Capturing schemes. 
After some algebraic pre-processing, we give explicit expressions
for the numerical viscosity terms
of two of the most widely used flux formulae, which implementation
saves computational time in multidimensional simulations of 
relativistic flows. Additionally, such treatment
explicitely cancells and factorizes a number of terms 
helping to amortiguate the growing of round-off errors.
We have checked the performance of our formulation running a 3D 
relativistic hydrodynamical code to solve a standard test-bed 
problem and found that the improvement in efficiency 
is of high practical interest in numerical simulations of 
relativistic flows in Astrophysics.
\end{abstract}

\thanks[thank]{This work has been supported by the Spanish DGES (grant
PB97-1432). The calculations were carried out on a SGI Origin 2000,
at the Centre de Inform\`atica de la Universitat de Val\`encia.}

\end{frontmatter}

\bf PACS \rm 47.11.+j, 47.75.+f, 95.30.Lz


Non-linear Systems of Conservation Laws.
High Resolution Shock Capturing methods.
Special Relativistic Hydrodynamics.
General Relativistic Hydrodynamics.


\newpage

The numerical study of the evolution of multidimensional 
relativistic flows turns out to be a topic of crucial interest 
in, at least, two different scientific fields: Nuclear Physics 
(studies of the properties of the equation of state for nuclear 
matter via comparison of simulations and experiments of heavy ion 
collisions) and Relativistic Astrophysics.
The field of Numerical Relativistic Astrophysics
is recently undergoing an extraordinary developement 
after the important efforts of people working in building up 
robust codes able to describe many different astrophysical scenarios,
such that relativistic jets in 
quasars and microquasars, accretion onto compact objects, collision 
of compact objects, stellar core collapse and recent models of 
Gamma-Ray bursts (see, e.g., the recent review 
in \cite{IM99} and references therein).
Thus, the improvement in
the efficiency of multidimensional hydro-codes becomes a necessity.

  It is well known the performance of {\it modern 
high-resolution shock-capturing} techniques (HRSC) 
in simulations of complex classical flows. 
Most of the HRSC methods are based on the solution 
of local Riemann problems
(i.e., initial value problems with discontiuous initial data) and
since 1991 \cite{MIM91} several Riemann Solvers or Flux Formulae 
have been specifically designed in relativistic
fluid dynamics (see, e.g., \cite{Mar97,IM99} for a review on 
Riemann solvers in Relativistic Astrophysics). 
In addition, in a recent paper \cite{PFIMM98} we showed
the way for applying special relativistic Riemann solvers in General Relativistic 
Hydrodynamics, hence any future new Riemann solver, exhaustively analyzed in Special 
Relativistic Hydrodynamics (SRH), could be applied to get the numerical 
solution of local Riemann problems in General Relativistic Hydrodynamics. 
Consequently, the interest of the results we obtain in this note goes 
beyond the domain of SRH and can be easily extended to General Relativistic 
Hydrodynamics. 

For consistency, we start by summarizing the basics of the HRSC 
techniques.
A system of {\it conservation laws} \cite{Le91} is
\be 
\frac{\partial {\bf u}}{\partial t}+
\sum_{i=1}^{3} \frac{\partial{\bf f}^{i}({\bf u})}
{\partial x_i} =0
\label{CLL}
\ee
\noindent
where {\bf u}$\in \Re^{d}$ is the {\it vector of unknowns} and
${\bf f}^{i}$({\bf u}) is the {\it flux} in the $i$-direction.
In the above system (\ref{CLL}) we can define a
$5\times 5$-Jacobian matrix ${\bf \cal B}^{i}$({\bf u}) associated 
to the flux in the $i$-direction as:
\begin{equation}
{\bf \cal B}^{i} = \frac{\partial{\bf f}^{i}({\bf u})}
{\partial\bf u \rm}.
\label{B}
\end{equation}
The system is said to be hyperbolic if the Jacobian matrices have
real eigenvalues.

The main ingredients of a HRSC algorithm are:

i) A finite discretization of the equations
in conservation form (\ref{CLL}).
Using a {\it method of lines}, this discretization reads:
\be
\frac{d{\bf u}_{i,j,k}(t) }{d t} =
-\frac{\hat{\bf f}_{i+{1\over 2},j,k}- \hat{\bf f}_{i-{1\over 2},j,k}}
{\Delta x} 
-\frac{\hat{\bf g}_{i,j+{1\over 2},k}- \hat{\bf g}_{i,j-{1\over 2},k}}
{\Delta y} 
-\frac{\hat{\bf h}_{i,j,k+{1\over 2}}- \hat{\bf h}_{i,j,k-{1\over 2}}}
{\Delta z}
\ee

\noindent
where subscripts $i, j, k$ are related, respectively, with $x$, $y$
and $z$-discretizations, and refer to cell-centered quantities. 
The cell width, in the three coordinate directions are, respectively,
$\Delta x$, $\Delta y$ and $\Delta z$. 

ii) Quantities $\hat{\bf f}_{i+{1\over 2},j,k}$,
$\hat{\bf g}_{i,j+{1\over 2},k}$ and
$\hat{\bf h}_{i,j,k+{1\over 2}}$ are called the {\it numerical fluxes} at 
the cell interfaces. These numerical fluxes are, in general, functions of
the states of the system at each side of the cell interface.  
Some HRSC methods derive expressions for the numerical fluxes by 
giving a consistent flux formulae or 
solving {\it local Riemann problems}, with an exact \cite{MM94}
or approximate Riemann solver, after a {\it cell reconstruction procedure}
that gives the state at both sides of the interface, denoted by $L$ 
(left state) and $R$ (right state). Several monotonic cell 
reconstruction prescriptions have been given in the scientific 
literature to achieve different orders of spatial accuracy 
\cite{VL79,wc,M94}. 
 
For clarity, from now on we will omit the indexes relative to the grid
and restrict our study to the $x_1$-splitting of 
the above system (\ref{CLL}), assuming that 
the vector of unknowns satisfies ${\bf u} = {\bf u} (x_1,t)$.

We have focussed our analysis to some of the most popular HRSC algorithms,
and analyzed their expressions for the numerical fluxes. Hence, the
sample considered is: HLLE \cite{HL83,E88},
Roe \cite{roe}, Marquina (M) \cite{DFIM98},
and a modified Marquina's flux formula (MM) \cite{AIMM99}. 
The above selection gathers the most fundamental differences among the large
sample of HRSC flux formulae. HLLE is the simplest one, it does not need
the full spectral decomposition of the Jacobian matrices. Roe's solver
linearizes the information contained in the spectral decomposition into 
an average state. Marquina's (and its modified version) flux formula
considers the information coming from each side of a given interface
(it is not a Riemann solver) and, in some astrophysical applications, 
has produced better results in modelling complex flows.

After some algebraic work, all these flux formulae can be 
cast into the following general form:
\be
\hat{\bf f}({\bf u}^L,{\bf u}^R) = \frac{1}{2}
\left(({\bf \cal I} + \widetilde{\bf \cal I}^L) {\bf f}^L + 
({\bf \cal I} - \widetilde{\bf \cal I}^R) {\bf f}^R  
 + ({\bf \cal Q}^L {\bf u}^L - {\bf \cal Q}^R {\bf u}^R)\right)
\label{FF}
\ee
\noindent
where ${\bf f}^{L,R}$ stands for the fluxes evaluated at the states
${\bf u}^{L,R}$ and ${\bf \cal I}$ is the unit matrix.
Following Harten \cite{Ha83},
the ${\bf \cal Q^{L,R}}$ terms in the above equation will be referred 
as the {\it numerical viscosity matrix}.

Matrices $\widetilde{\bf \cal I}^{L,R}$ and ${\bf \cal Q}^{L,R}$ 
can be expressed as linear combinations of the projectors onto each
eigenspace, i.e., the direct product of the corresponding left and right 
eigenvectors ${\bf l}_p, {\bf r}_p$ associated to the p-th characteristic
field (p=1,...,d),

\be
\widetilde{\bf \cal I}^{L,R} = \displaystyle{\sum_{p=1}^d b_p {\bf l}_p^{L,R} 
{\bf r}_p^{L,R}}
\ee
\be
{\bf \cal Q}^{L,R} = \displaystyle{\sum_{p=1}^d c_p {\bf l}_p^{L,R} 
{\bf r}_p^{L,R}} 
\label{matrices} 
\ee
where superscripts $L,R$ indicate that the eigenvectors are evaluated 
at the state ${\bf u}^{L,R}$.
The values of the coefficients $b_p$ and $c_p$ appearing in the above
definitions of matrices $\widetilde{\bf \cal I}^{L,R}$ and ${\bf \cal Q}^{L,R}$ 
depend on the eigenvalues $\lambda_p$ as shown in Table I, for the four
flux formulae analyzed. 

\begin{table}
    
\centerline{\begin{tabular}{|ccccc|}
    \multicolumn{5}{c}{\bf TABLE I \rm} \\
    \multicolumn{5}{c}{\protect \small Parameters in the numerical 
fluxes.} \\
\hline 
\multicolumn{5}{|c|}{} \\
Flux &&  $b_p$  && $ c_p $  \\
Formulae &&   &&   \\
\multicolumn{5}{|c|}{} \\
\hline 
\multicolumn{5}{|c|}{} \\
HLLE  &&  $\displaystyle{\frac{\Psi_{+} + \Psi_{-}}{\Psi_{+} - \Psi_{-}} }$
        && $ \frac{1}{2}(\Psi_{+} - \Psi_{-}) $ \\ 
Roe   &&  0   && $\mid \lambda_p(\widetilde{\bf u}) \mid $ \\ 
M && $\beta_p $ && $ \alpha_p (1-\beta_p^2)$ \\ 
MM && 0  && $\alpha_p $ \\ 
\hline
\end{tabular}}
\caption{In the above table we have introduced the quantities
$\Psi_{+} = max (0, \lambda_{+}^R, \lambda_{+}^L)$ and
$\Psi_{-} = min (0, \lambda_{-}^R, \lambda_{-}^L)$,
$\lambda_{+}$ and $\lambda_{-}$ are, respectively, the maximum and
minimum of $\lambda_{p}$,  
$\alpha_p = max \left(\mid \lambda_p^L \mid, \mid \lambda_p^R \mid \right)$ 
and $\beta_p = \frac{1}{2} \left(sgn(\lambda_p^L)+ sgn(\lambda_p^R)\right)$.
We denote by $\widetilde{\bf u}$ the state of the system 
according to Roe's average.}

\end{table}

Several comments concerning Table I are in order: 

i) If we take into account the orthonormality relations between the
right and left eigenvectors
\be
\displaystyle{\sum_{p=1}^d} {\bf l}_p {\bf r}_p = {\bf \cal I}
\ee
and the fact that the coefficients 
$b_p$ and $ c_p $ are, in the case of HLLE, independents of $p$, 
then matrices $\widetilde{\bf \cal I}^{L,R}$ and ${\bf \cal Q}^{L,R}$ are,
trivially, the unit matrix multiplied by the corresponding factors.

ii) For HLLE's and Roe's flux formulae their corresponding matrices 
$\widetilde{\bf \cal I}^{L,R}$ and ${\bf \cal Q}^{L,R}$ satisfy the relations: 
$\widetilde{\bf \cal I}^{L}$ = $\widetilde{\bf \cal I}^{R}$,
${\bf \cal Q}^{L}$ = ${\bf \cal Q}^{R}$

iii) As it is well known, the knowledge of the spectral decomposition
of the Jacobian matrices is a basic ingredient to build up Riemann 
solvers or many flux formulae. Nevertheless, while HLLE's flux formula 
only needs the values of the maximum and minimum speeds of
propagation of the signals, Roe's and Marquina's flux formulae 
need explicitly the full knowledge of the spectral decomposition,
including right and left eigenvectors. 

The system governing the evolution of multidimensional relativistic 
perfect fluids can be written in Cartesian coordinates
in the form (\ref{CLL}), with $d=5$,
where, in units such that the speed of light $c=1$, 
the vector of unknowns ${\bf u}$ is given by
\begin{equation}
{\bf u} = \left(D, \: S^1, \: S^2, \: S^3, \: {\tau} \right)^{T} \, ,
\label{U}
\end{equation}
the fluxes are defined by
\be
{\bf f}^{i} = \left(D v^{i}, S^1 v^i + p \delta^{1i},
S^2 v^i + p \delta^{2i}, S^3 v^i + p \delta^{3i}, S^i - Dv^i \right)^T 
\ee
\noindent
where $D (= \rho W) $ is the rest mass density, $S^{j} (= \rho h W^2 v^j)$ is 
the j-component of the momentum density, and $\tau (= \rho h W^{2}-p-\rho W)$ 
is the energy density, $W = (1-{\rm v}^{2})^{-1/2}$ is
the Lorentz factor,
$\rho$ is the rest--mass density, $p$ the pressure and
$h$ the specific enthalpy given by $h = 1 + \varepsilon + p/\rho$ with
$\varepsilon$ being the specific internal energy. 
The system of SRH 
is closed with an equation of state $p=p(\rho,\varepsilon)$ from which 
the local sound speed, $c_s$, can be obtained
\begin{equation}
 h c_{s}^{2} = \frac{\partial p }{ \partial \rho} + (p/\rho^{2})
 \frac{\partial p }{ \partial \epsilon},
 \label{cs}
\end{equation}

In a previous paper \cite{DFIM98} we derived 
the explicit analytical expressions for
the full (right and left) spectral decomposition. We denote  
the five characteristic fields by $p= {1,\ldots,5} 
\equiv {-,0,0,0,+}$, in the standard ordenation. 

A very worthy simplification on the calculation of matrices 
${\bf \cal Q}$ 
arises when some eigenvalue is degenerate, i.e., when the system
is not strictly hyperbolic. In SRH, like in multidimensional Newtonian 
hydrodynamics, there is a {\it linearly degenerate} field, $p=0$, such 
that the corresponding eigenvalue $\lambda_0$ is triple 
(the system in the $j$-direction splitting is not strictly hyperbolic, 
although the set of eigenvectors is complete). According to equation 
(\ref{matrices}), and using the orthonormality relations between the 
right and left eigenvectors 
\be
\sum_{k=1}^{3} r^m_{0,k} l^n_{0,k} = \delta^{mn} - r^m_+ l^n_+ - r^m_- l^n_-
\ee
where $m,n=1, \ldots 5$ denote the components of the 5-vector,
it is possible to eliminate the three eigenvectors associated to the
degenerate field and to write down the following simplified 
expression (omitting $L$,$R$ superscripts) 
\be
{\bf \cal Q}^{mn} =  c_0 \delta^{mn} + (c_+ - c_0) r^m_+ l^n_+
+ (c_- - c_0) r^m_- l^n_- . 
\ee
 
Notice that only $r_{\pm}$ and $l_{\pm}$
are needed to evaluate the numerical viscosity. The same procedure 
can be applied to any system of conservation laws where one of the 
eigenvectors has multiple degeneracy, because orthogonality relations 
always allow us to skip the explicit dependence on one of the 
vector subspaces of the spectral decomposition. In particular, 
it could be of great interest in the case of the
equations of relativistic magnetohydrodynamics where,
in the ansatz of a directional splitting, similar
degeneracy arises in the structure of the characteristic fields
associated to each one of the fluxes.

The explicit formulae for the numerical viscosity term corresponding 
to the system of equations of special relativistic hydrodynamics are:

{\it HLLE's flux formulae}. Since the numerical viscosity matrix is
proportional to the identity, the application of the above recipes 
is obvious.

{\it Roe's flux formulae}. The numerical flux across some given interface can be written
\be
\hat{\bf f}({\bf u}^L,{\bf u}^R) = \frac{1}{2}[{\bf f}^L +{\bf f}^R + {\bf q}]
\label{numfluxroe}
\ee
\noindent
${\bf q}$ being the five--vector calculated from the corresponding numerical
viscosity matrices of Table I:
\be
{\bf q} = {\bf \cal Q} ({\bf u}^L-{\bf u}^R) 
\equiv {\bf \cal Q} \Delta {\bf u} 
\ee

In Roe's Riemann solver the quantities relative to the spectral decomposition are
evaluated using the corresponding Roe-averages of the left and right states,
denoted by $\widetilde{\bf u}$ 
(see \cite{roe}, for the Newtonian case and \cite{EM95} 
for the relativistic case). 
In practice, other particular averaging (e.g., arithmetic means) 
have also been used. Note that
in the following expressions (\ref{qroe}) all quantities are
evaluated using Roe's average, except for $\Delta u_m$.
After some algebra, the viscosity vector associated to the numerical
flux in the $j$-direction is

\begin{eqnarray}
q_1 & = & \mid {\lambda_0} \mid \Delta u_1 + \chi_a 
\nonumber \\
q_2 & = & \mid {\lambda_0} \mid \Delta {u}_2 + h W (v_x \chi_a + \chi_b \delta_{jx}) 
\nonumber \\
q_3 & = & \mid {\lambda_0} \mid \Delta {u}_3 + h W (v_y \chi_a + \chi_b \delta_{jy})
\nonumber \\
q_4 & = & \mid {\lambda_0} \mid \Delta {u}_4 + h W (v_z \chi_a + \chi_b \delta_{jz}) 
\nonumber \\
q_5 & = & \mid {\lambda_0} \mid \Delta {u}_5 + h W (\chi_a + v_j \chi_b) 
- \chi_a 
\label{qroe}
\end{eqnarray}
where
\begin{eqnarray}
\chi_a & = &  \sum_{m=1}^5 \left[(\mid {\lambda_+} \mid - \mid {\lambda_0} 
\mid) l^m_+ + (\mid {\lambda_-} \mid - \mid {\lambda_0} \mid) l^m_- \right] 
\Delta u_m \\
\chi_b & = &  \sum_{m=1}^5 
\left[(\mid {\lambda_+} \mid - \mid {\lambda_0} \mid) V^j_+ l^m_+
+ (\mid {\lambda_-} \mid - \mid {\lambda_0} \mid) V^j_- l^m_- \right] 
\Delta u_m \\
V^j_{\pm} &  = & {\displaystyle{\frac{\lambda_{\pm} - v^j}{1-v^j \lambda_{\pm}}}} 
\end{eqnarray}


{\it M and  MM- flux formulae}. The numerical flux across 
a given interface can be 
written like equation (\ref{numfluxroe}) with
\be
{\bf q} = {\bf q}^{L} - {\bf q}^{R} 
\ee
\be
{\bf q}^{L,R} = {\bf \cal Q}^{L,R} {\bf u}^{L,R} 
\ee
Omitting the superscripts $L,R$ and taken into account
the expressions in Table I for MM and the results in
\cite{DFIM98}, the viscosity vector in the $x$-splitting is:
\begin{eqnarray}
q^{L,R}_1 &=& \frac{h^2}{\Delta} \left\{ M \left[ {\cal A}_{-} {\Omega}_{+} - 
                                          {\cal A}_{+} {\Omega}_{-} \right] + 
p(c_{+}\aleph_{+}-c_{-}\aleph_{-}) \right\} + 
                                \nonumber \\ 
& & c_{0}p\frac{W}{h}\left\{ \frac{{\cal K}}{{\cal K} - 1} + 
\frac{v_y^2+v_{z}^2} {1-v_x^2} \right \} 
\nonumber \\
q^{L,R}_2 &=& \frac{h^2W}{\Delta} \left\{ M{\cal A}_{+}{\cal A}_{-} \left[ 
{\Omega}_{+}\lambda_{+}-{\Omega}_{-}\lambda_{-}
\right] + 
p(c_{+}\lambda_{+}{\cal A}_{+}\aleph_{+}-
         c_{-}\lambda_{-}{\cal A}_{-}\aleph_{-}) \right\} + 
                                \nonumber \\ 
& & c_{0}pW^2v_x \left\{ \frac{1}{{\cal K} - 1} + 
2\frac{v_y^2+v_{z}^2}{1-v_x^2} 
\right \} 
\nonumber \\
q^{L,R}_3 &=& \frac{h^2W}{\Delta}v_y \left\{ M \left[ 
{\Omega}_{+}{\cal A}_{-} -
{\Omega}_{-}{\cal A}_{+} \right] + 
p(c_{+}\aleph_{+}-c_{-}\aleph_{-}) \right\} + 
                                \nonumber \\ 
& & c_{0}p \left\{ \frac{W^2}{{\cal K} - 1} + 
\frac{1+2W^2(v_y^2+v_{z}^2)}{1-v_x^2} 
\right \} 
\nonumber \\
q^{L,R}_4 &=& \frac{h^2W}{\Delta}v_{z} \left\{ M \left[ 
{\Omega}_{+}{\cal A}_{-}-
{\Omega}_{-}{\cal A}_{+} \right] + 
p(c_{+}\aleph_{+}-c_{-}\aleph_{-}) \right\} + 
                                \nonumber \\ 
& & c_{0}p \left\{ \frac{W^2}{{\cal K} - 1} + 
\frac{1+2W^2(v_y^2+v_{z}^2)}{1-v_x^2} 
\right \} 
\nonumber \\
q^{L,R}_5 &=& \frac{h^2}{\Delta} \left\{ M \left[ {\cal A}_{-} {\Omega}_{+}{\cal D}_{+}-
                     {\cal A}_{+} {\Omega}_{-}{\cal D}_{-} \right] + 
p[c_{+}\aleph_{+}{\cal D}_{+} -
         c_{-}\aleph_{-}{\cal D}_{-}] \right\} + 
                                \nonumber \\ 
& & c_{0}p\frac{W}{h}\left\{ \frac{hW-{\cal K}}{{\cal K} - 1} + 
\frac{(2hW-1)(v_y^2+v_{z}^2)}{1-v_x^2} \right \},
\label{viscos}
\end{eqnarray}
with the following auxiliary quantities

\begin{eqnarray}
M=\rho hW^2({\cal K}-1), \quad
{\Omega}_{\pm}= c_{\pm}(v_x - \lambda_{\mp}), \quad
{\cal D}_{\pm}= hW{\cal A}_{\pm}-1, 
\nonumber \\
 {\cal K} \equiv
 {\displaystyle{\frac{\tilde{\kappa}}
 {\tilde{\kappa}-c_s^2}}, \:\:\:\:
 {\tilde \kappa}= \frac{1}{\rho}
 \left.\frac{\partial p}{\partial \varepsilon}\right|_{\rho}},
 \:\:\:\:
 {\cal A}_{\pm} \equiv
 {\displaystyle{\frac{1 - v_x v_x}{1 - v_x {\lambda}_{\pm}}}}
\nonumber\\
\Delta = h^3 W ({\cal K} - 1) (1 - v_x v_x)
({\cal A}_{+} {\lambda}_{+} - {\cal A}_{-} {\lambda}_{-})
\nonumber \\
\aleph_{\pm} = 
{\pm}\left \{ -v_x -W^2({\rm v}^{2}-v_x v_x)(2{\cal{K}}-1)(v_x - 
{\cal{A}_{\pm}}\lambda_{\pm})
+{\cal{K} \cal{A}_{\pm}} \lambda_{\pm} \right\}
\end{eqnarray}
where quantities $c_{\pm,0}$ are given in Table I and $\Delta$ is 
the determinant of the matrix of right-eigenvectors.

The corresponding viscosity vectors in the other directions are trivially
obtained by a cyclic permutation of subindices ${x,y,z}$.

We have tested the efficiency of our numerical proposal, 
for Roe's and MM's flux formulae,
running GENESIS (a 3D special relativistic hydro-code  \cite{AIMM99}),
without any optimization at compilation level, in a SGI Origin 2000.
A standard initial value problem has been chosen:
$\rho_L=10$, $\epsilon_L=2$, $v_L=0$, $\gamma_L=5/3$, $\rho_R=1$,
$\epsilon_R=10^{-6}$, $v_R=0$ and $\gamma_R=5/3$, where the subscript
$L$ ($R$) denotes the state to the left (right) of the initial
discontinuity. This test problem has been considered by several
authors in the past (see \cite{AIMM99} for details in 1D, 2D and 3D).

We have compared the performance of two different implementations of the
numerical flux subroutine:

        i) {\it Case A}, stands for the results obtained using our analytical
prescription. This means to write down, in the numerical flux routine, just the 
expressions derived here for the viscosity vector ${\bf q}$.

        ii) {\it Case B}, stands for the results obtained running the code with a 
standard high-efficiency subroutine for inverting matrices (we
use a LU decomposition plus an implicit pivoting which is, for general
matrices, $O(N^3)$). This subroutine is called to 
get the left eigenvectors from the matrix of right eigenvectors 
and is adapted to the particular dimensions of the matrices
($3 \times 3$, in 1D, $4 \times 4$, in 2D and $5 \times 5$, in 3D).
Hence, unlike case A, now we have to calculate numerically the
following quantities: the matrix of left eigenvectors, the 
characteristic variables and, finally, the components of the viscosity 
vector ${\bf q}$.

\begin{table}
\centerline{\begin{tabular}{|cccccc|}
    \multicolumn{6}{c}{\bf TABLE II \rm} \\
    \multicolumn{6}{c}{\protect \small CPU time (in microseconds).} \\
    \multicolumn{6}{c}{}\\
    \hline
    \hline
     &   & \multicolumn{4}{c|}{TCI $(\mu s)$}  \\
     &   & \multicolumn{2}{c}{Roe} & \multicolumn{2}{c|}{MM}  \\
    Case &  \# Zones  & Case $A$  & Case $B$ & Case $A$  & Case $B$ \\
    \hline
    1D  & $100 \times 1 \times 1 $ & $12.2$ & $53.8$ & $23.8$ & $118.9$   \\
    2D  & $20 \times 20 \times 1 $ & $25.5$ & $181.8$ & $49.0$ & $373.5$  \\
    3D  & $14 \times 14 \times 14$ & $39.4$ & $431.9$ & $75.7$ & $879.0$  \\
\hline
\hline 
\end{tabular}}
\caption{\small Time per numerical cell 
and iteration (TCI) in microseconds employed by the numerical 
flux routine in our test-bed problem, for three different grids.}
\end{table}

Table II summarizes the results: 
the direct implementation of our numerical viscosity formulae
leads to an improvement of the efficiency (in terms of CPU time)
of the numerical fluxes subroutine
in a factor which, in 3D calculations, ranges between about {\it eleven} and 
{\it twelve} depending on the particular flux formula used.
When comparing Roe's and MM's cases
a factor two --in favour of Roe-- arises due to the fact that MM's flux
formulae needs to compute two viscosity vectors (one per each side of a given
interface), unlike Roe's flux formula which needs only one viscosity 
vector evaluated at the average state. As it must be, the efficiency 
increases with the number of spatial dimensions involved in the 
problem due to the computationally expensive matrix inverting 
operations performed at each interface to get the numerical fluxes. 
Since the numerical flux routine is, typically, one of the 
most time-consuming, it translates into a speed up factor
between {\it two} and {\it four} in
the total execution time, depending on the specific weight of the 
flux formulae routine in each particular application.

 Our formulation also gives a unified description of the numerical
fluxes (\ref{FF}), permitting a unique 
implementation with the possibility of switching in cases when the 
utilisation of a specific flux formula is more appropriate. 
In addition, due to the fact that we have eliminated, in the 
generalized MM's flux formula, all the {\it conditional clauses},
the efficiency is ensured either for scalar or vectorial processors.

  Another worthy by-products of our algebraic pre-processing concerns
with the significant reduction of round-off errors, 
as a consequence of the number of operations suppressed and 
factorization. 
One of the important issues in designing a multidimensional
hydro-code is the accurate preservation of 
any symmetries present in a physical problem.
A numerical violation of these symmetries could arise
as a consequence of accumulation of round-off errors in the
calculation of the numerical fluxes, as
we have explained in a previous paper \cite{AIMM99}.
The algebraic simplifications, shown in the present paper, 
reduce the number of operations and cure such problem.
 
    Two last additional consequences arise from our work. First is that 
similar expressions can be worked out for any non-linear hyperbolic 
system of conservation laws for which the full spectral decomposition 
is known. In particular, when some of the vectorial subspaces has
multiple degeneracy, a similar algebraic preprocessing is very convenient.  
The other important consequence is that an appropriate combination of 
a simplified formulation of the numerical viscosity together with 
the use of special relativistic Riemann solvers in General 
Relativistic Hydrodynamics \cite{PFIMM98}, should allow a very 
easy and efficient extension to General Relativistic Hydrodynamics.


\end{document}